\documentstyle[amssymb,preprint,aps]{revtex}

\begin{document}
\title{Dynamics of spin-2 Bose condensate driven by external magnetic fields}
\author{Ping Zhang$^{1,2}$, C. K. Chan$^3$, Xiang-Gui Li$^4$, Qi-Kun Xue$^1$,
Xian-Geng Zhao$^2$}
\address{$^1$International Center of Quantum Structure and State Key Laboratory for
Surface Physics, Institute of Physics, The Chinese Academy of Sciences,
Beijing 100080, P.R. China\\
$^2$Institute of Applied Physics and Computational Mathematics, P.O.Box
8009, Beijing 100088\\
$^3$Department of Applied Mathematics, The Hong Kong Polytechnic University,
Hong Kong\\
$^4$Department of Applied Mathematics, University of Petroneum, Shandong
257062, China}
\maketitle

\begin{abstract}
Dynamic response of the $F=2$ spinor Bose-Einstein condensate (BEC) under
the influence of external magnetic fields is studied. A general formula is
given for the oscillation period to describe population transfer from the
initial polar state to other spin states. We show that when the frequency
and the reduced amplitude of the longitudinal magnetic field are related in
a specific manner, the population of the initial spin-0 state will be
dynamically localized during time evolution. The effects of external noise
and nonlinear spin exchange interaction on the dynamics of the spinor BEC
are studied. We show that while the external noise may eventually destroy
the Rabi oscillations and dynamic spin localization, these coherent
phenomena are robust against the nonlinear atomic interaction.%
\newline%
PACS number(s): 03.75.Fi, 05.30.Jp, 32.80.Pj%
\newline%
\end{abstract}

\section{INTRODUCTION}

Recent experiments on $^{23}$Na condensates confined in an optical trap\cite
{ref1,ref2,ref3} have stimulated extensive interest in the study of
multicomponent spinor Bose-Einstein condensates (BECs). Due to the hyperfine
spin of atoms, BECs of alkali-metal atoms have internal degrees of freedom
which are frozen in a magnetic trap. Introduction of an optical trap
liberates them to allow BEC to be in a superposition of magnetic sublevels.
Therefore, the spinor BEC can be represented by a vector order parameter.
The $F=1$ spinor BEC was first theoretically studied by Ho\cite{ref4}, Ohmi
and Machida\cite{ref5} by generalizing the Gross-Pitaevskii equation under
the restriction of gauge and spin-rotation symmetry. Within the mean-field
theory, they predicted a rich set of new phenomena such as spin textures and
topological excitations. Law {\it et al.}\cite{ref6} constructed an
excellent algebraic representation of the $F=1$ BEC Hamiltonian to study the
exact many-body states, and found that spin-exchange interactions cause a
set of collective dynamic behavior of BEC. Since the spinor BEC appears
feasible by using the $F=2$ multiplet of bosons such as $^{23}$Na, $^{87}$%
Rb, or $^{85}$Rb, it is necessary to investigate the ground-state structure
and magnetic response of $F=2$ spinor BEC. Recently, Ciobanu {\it et al}. 
\cite{ref7} generalized the approach for the $F=1$ spinor BEC developed by
Ho to study the ground-state structure of the $F=2$ spinor BEC. They found
that there are three possible phases in zero magnetic field, which are
characterized by a pair of parameters describing the ferromagnetic order and
the formation of singlet pairs. From current estimates of scattering lengths
they also found that the spinor BEC of $^{87}$Rb and $^{23}$Na have a polar
ground state, whereas those of $^{85}$Rb and $^{83}$Rb are cyclic and
ferromagnetic, respectively. Koashi and Ueda\cite{ref8} studied the exact
eigenspectra and eigenstates of $F=2$ spinor BEC. They found that, compared
to $F=1$ spinor BEC, the $F=2$ spinor BEC exhibits an even richer magnetic
response due to quantum correlations among three bosons.

Different from superfluid $^3$He, a new feature in a spinor BEC is that its
response to an external magnetic field is dominated by the electronic rather
than nuclear spin. This opens up possibilities of manipulating the magnetism
of superfluid vapors. Observation of spin domains by Stenger {\it et al}. at
MIT\cite{ref2} offers a remarkable example of such manipulations. Pu {\it et
al.}\cite{ref9}{\it \ }investigated the effects of external magnetic fields
on the dynamics of $F=1$ spinor BEC and found various magnetic field-induced
effects, such as stochastization in population evolution, metastability in
spin composition, and dynamic localization in spin space. In this paper we
investigate the time evolution of the $F=2$ spinor BEC in the presence of
external magnetic fields with longitudinal and transverse field strengths $%
B_z(t)$ and $B_x$, respectively. Here the longitudinal magnetic field lifts
the energy degeneracy of the spin states through Zeeman effects, whereas the
transverse field appears as a coupling between different spin components. We
assume the longitudinal field to be time-dependent and that the transverse
field is static. Using perturbation theory, a general formula for the
oscillation period to describe the passing across the initial spin-polarized
state is obtained. Another aspect we are interested is the effects of the
external noise and the nonlinear spin-exchange interactions on the quantum
coherent behavior of the spinor BEC.

In Sec. II the Hamiltonian for the spin-2 BEC system is presented. The Rabi
oscillation and dynamic spin localization is shown in Sec. III. In Sec. IV
we discuss the effects of external noise on the dynamics of the spinor BEC.
A full numerical discussion based on an effective one-dimensional nonlinear
Schr\"{o}dinger equation is given in Sec. V. A summary is given in Sec. VI.
\ 

\section{THE MODEL: Single-mode approximation}

We consider the $F=2$ spinor BEC subject to a spatial-uniform magnetic field 
$\widehat{B}(t)$. The Hamiltonian invariant under spin space rotation and
gauge transformation is written in terms of the five-component field
operators: $\Psi _{+2},...,\Psi _{-2},$ corresponding to the sublevels $%
m_F=+2,...,-2$ of the hyperfine state $F=2$. Namely, it is given\cite
{ref7,ref8} by $H=H_0+H_B$,

\begin{eqnarray}
H_0=\int d{\bf r}[\frac{\hslash ^2}{2M}\nabla \Psi _\alpha ^{+}\cdot \nabla
\Psi _\alpha +U\Psi _\alpha ^{+}\Psi _\alpha +\frac{\overline{c}_0}2\Psi
_\alpha ^{+}\Psi _\beta ^{+}\Psi _\beta \Psi _\alpha &&+\frac{\overline{c}_1}%
2\sum_i(\Psi _\alpha ^{+}(F_i)_{\alpha ,\beta }\Psi _\beta )^2  \nonumber \\
+\overline{c}_2\Psi _\alpha ^{+}\Psi _{\alpha ^{\prime }}^{+}\langle 2\alpha
;2\alpha ^{\prime } &\mid &00\rangle \times \langle 00\mid 2\beta ;2\beta
^{\prime }\rangle \Psi _\beta \Psi _{\beta ^{\prime }}],
\end{eqnarray}

\begin{equation}
H_B=-\mu _Bg_f\int d{\bf r}\Psi _\alpha ^{+}(\widehat{B}(t)\cdot F)_{\alpha
,\beta }\Psi _\beta ,
\end{equation}
where $\overline{c}_0,$ $\overline{c}_1$, and $\overline{c}_2$ are related
to scattering lengths $a_0,$ $a_2,$ and $a_4$ of the two colliding bosons,
with total angular momenta 0, 2, and 4, by $\overline{c}_0=4\pi \hslash
^2(3a_4+4a_2)/7M,$ $\overline{c}_1=4\pi \hslash ^2(a_4-a_2)/7M,$ and $%
\overline{c}_2=4\pi \hslash ^2(3a_4-10a_2+7a_0)/7M$. In addition, we have
introduced in Eqs. (1) and (2) $5\times 5$ spin matrices $F_i$ ($i=x,y,z$)

\begin{equation}
F_{x}=\left( 
\begin{array}{ccccc}
0 & 1 & 0 & 0 & 0 \\ 
1 & 0 & \sqrt{3/2} & 0 & 0 \\ 
0 & \sqrt{3/2} & 0 & \sqrt{3/2} & 0 \\ 
0 & 0 & \sqrt{3/2} & 0 & 1 \\ 
0 & 0 & 0 & 1 & 0
\end{array}
\right) ,
\end{equation}

\begin{equation}
F_{y}=-i\left( 
\begin{array}{ccccc}
0 & 1 & 0 & 0 & 0 \\ 
-1 & 0 & \sqrt{3/2} & 0 & 0 \\ 
0 & -\sqrt{3/2} & 0 & \sqrt{3/2} & 0 \\ 
0 & 0 & -\sqrt{3/2} & 0 & 1 \\ 
0 & 0 & 0 & -1 & 0
\end{array}
\right) ,
\end{equation}

\begin{equation}
F_z=\left( 
\begin{array}{ccccc}
2 & 0 & 0 & 0 & 0 \\ 
0 & 1 & 0 & 0 & 0 \\ 
0 & 0 & 0 & 0 & 0 \\ 
0 & 0 & 0 & -1 & 0 \\ 
0 & 0 & 0 & 0 & -2
\end{array}
\right) .
\end{equation}
We assume the external magnetic field is weak so that the coordinate wave
function $\Phi ({\bf r})$ is independent of the spin state and solely
determined by the first three terms of Eq. (1), namely, $[-\hslash ^2\nabla
^2/2M+U+\overline{c}_0(N-1)\left| \Phi \right| ^2]\Phi =\mu \Phi $, with $N$
being the total particle number. Substituting $\Psi _\alpha =\widehat{a}%
_\alpha \Phi $ into Eqs. (1) and (2) and keeping only spin-dependent terms,
we obtain

\begin{equation}
H_{0}^{\prime }=\frac{c_{1}}{2}\widehat{F}\cdot \widehat{F}+\frac{2c_{2}}{5}%
\widehat{S}_{+}\widehat{S}_{-},
\end{equation}
and

\begin{equation}
H_B=-\mu _Bg_f\widehat{B}(t)\cdot \widehat{F},
\end{equation}
where $\widehat{F}_i=\widehat{a}_\alpha ^{+}(F_i)_{\alpha \beta }\widehat{a}%
_\beta $, $\widehat{S}_{+}=\widehat{S}_{-}^{+}=(\widehat{a}_0^{+})^2/2-%
\widehat{a}_1^{+}\widehat{a}_{-1}^{+}+\widehat{a}_2^{+}\widehat{a}_2^{+}$,
and $c_i=(\overline{c}_i)\int d{\bf r}\left| \Phi \right| ^4$. We further
assume that the magnetic fields consist of longitudinal and transverse
components. Without loss of generality, the transverse direction of the
field is chosen to be along the $x$ axis, i.e., $\widehat{B}(t)=B_l(t)%
\widehat{z}+B_x\widehat{x}$. In such a case, the second-quantized
Hamiltonian of $H_B$ is

\begin{eqnarray}
H_B &=&-\mu _Bg_fB_l(t)(\widehat{a}_2^{+}\widehat{a}_2+\widehat{a}_1^{+}%
\widehat{a}_1-\widehat{a}_{-1}^{+}\widehat{a}_{-1}-\widehat{a}_{-2}^{+}%
\widehat{a}_{-2})  \nonumber \\
&&-\mu _Bg_fB_x(\widehat{a}_2^{+}\widehat{a}_1+\sqrt{3/2}\widehat{a}_1^{+}%
\widehat{a}_0+\sqrt{3/2}\widehat{a}_0^{+}\widehat{a}_{-1}+\widehat{a}%
_{-1}^{+}\widehat{a}_{-2}+h.c.).
\end{eqnarray}
{\it \ } Similar to Ciobanu {\it et al.}\cite{ref7}, mean-field
approximation is used such that the field operators $\widehat{a}_\alpha $
are replaced by $c$ numbers $a_\alpha =\sqrt{P_\alpha }e^{i\theta _\alpha }$%
, where $P_\alpha =N_\alpha /N$ is the population in spin $\alpha $ and $%
\theta _\alpha $ the phase of wave function $a_\alpha $. Furthermore, since
this paper deals with the quantum coherent behavior of the system under the
influence of the external magnetic fields, we assume that the initial spin
state of the BEC is the eigenstate in the absence of external fields,
consequently, contribution from Hamiltonian (6) is a constant energy shift,
and can be neglected in the dynamics. The semiclassical equations of motion
in Heisenberg representation can be derived from the Hamiltonian $H_B$ (We
introduce the state vector $a=(a_2,...,a_{-2})^T$)

\begin{equation}
i\stackrel{.}{a}=H_{eff}(t)a,
\end{equation}
where 
\begin{equation}
H_{eff}(t)=-b_{l}(t)F_{z}-b_{x}F_{x}.
\end{equation}
Here we have defined $b_{l}(t)=\mu _{B}g_{f}B_{l}(t)$ and $b_{x}=\mu
_{B}g_{f}B_{x}$. The time evolution of the system has a rich set of
intriguing features with different initial conditions. In this paper, we
consider the case that the system begins with unperturbed spin-0 state $%
a(0)=(0,0,1,0,0)^{T}$, i.e., the $P0$ polar phase defined by Ciobanu {\it et
al.}\cite{ref7}. This phase is possible to be realized in $^{87}$Rb and $%
^{23}$Na.

\section{RABI OSCILLATIONS AND DYNAMIC\ SPIN\ LOCALIZATION}

In this section, we analyze the time evolution of $F=2$ BEC described by Eq.
(9) with initial polar state $a(0)=(0,0,1,0,0)^{T}$. Assuming that $%
a(t)=e^{i\int_{0}^{t}d\tau b_{l}(\tau )F_{z}}\varphi (t)$, we obtain

\begin{equation}
i\stackrel{.}{\varphi }(t)=H_{I}(t)\varphi (t),
\end{equation}
where $H_{I}(t)=-b_{x}\exp (-i\int_{0}^{t}d\tau b_{l}(\tau )F_{z})F_{x}\exp
(i\int_{0}^{t}d\tau b_{l}(\tau )F_{z})$. Making use of the identity

\begin{equation}
\exp (i\lambda F_{z})F_{x}\exp (-i\lambda F_{z})=F_{x}\cos \lambda
-F_{y}\sin \lambda ,
\end{equation}
the formal solution of Eq. (11) is obtained

\begin{equation}
\varphi (t)=\widehat{T}e^{i\int_{0}^{t}d\tau \lbrack X(\tau )F_{x}+Y(\tau
)F_{y}]}\varphi (0),
\end{equation}
where $X(t)=b_{x}\cos [\int_{0}^{t}d\tau b_{l}(\tau )]$, $Y(t)=b_{x}\sin
[\int_{0}^{t}d\tau b_{l}(\tau )]$, and $\widehat{T}$ denotes time ordering.
Considering the situation in which the transverse magnetic field $b_{x}$ is
weak, we may approximate Eq. (13) by the following solution

\begin{equation}
\varphi (t)=e^{i\int_{0}^{t}d\tau \lbrack X(\tau )F_{x}+Y(\tau
)F_{y}]}\varphi (0).
\end{equation}
This solution is valid to the first order in $b_{x}$ and preserves
unitarity. After a straightforward calculation, we obtain the time evolution
of the population of spin-$\alpha $ state as

\begin{equation}
P_\alpha (t)=\left| a_\alpha (t)\right| ^2=\left| \sum_{\beta =-2}^2d_{\beta
\alpha }^2(\pi /2)d_{\beta 0}^2(\pi /2)e^{-i\alpha \int_0^td\tau b_l(\tau
)}e^{i\beta \lambda }\right| ^2,
\end{equation}
where $\lambda =\sqrt{(\int_0^td\tau X(\tau ))^2+(\int_0^td\tau Y(\tau ))^2}$%
, and the matrix element is defined as

\begin{equation}
d_{\beta \alpha }^2(\theta )=\langle 2\beta \mid e^{-i\theta F_y}\mid
2\alpha \rangle ,\text{ }(\beta ,\alpha =2,...,-2).
\end{equation}
In particular, the population of the initial spin-0 state at time $t$ is

\begin{equation}
P_0(t)=\left| a_0(t)\right| ^2=\cos ^2\left| \int_0^td\tau
_1b_xe^{i\int_0^{\tau _1}d\tau _2b_l(\tau _2)}\right| .
\end{equation}
Thus to the first-order approximation in $b_x$ we obtain the analytical
expression of the evolution probability for the spinor BEC remaining in the
initial polar state. Expression (17) is valid for arbitrary time-dependent
external magnetic fields. Obviously, $e^{i\int_0^td\tau b_l(\tau )}$ can be
expanded as a discrete Fourier series, and the time integral is either
bounded or increases linearly in time (on top of an oscillatory piece). In
the former case, $P_0$ remains close to unity at all times because of the
smallness of $b_x$. This high population of the initial spin state is just
the spin localization. In the latter case, $P_0$ oscillates between $0$ and $%
1$, implying population transfer between $P_0$ and other $P_\alpha $ ($%
\alpha =\pm 1$, $\pm 2$). Hence we have the spin delocalization, i.e., Rabi
oscillation. Thus, we obtain the oscillation period $T$ as

\begin{equation}
T=\frac{\pi }{\lim_{t\rightarrow \infty }\left| \frac{1}{t}\int_{0}^{t}d\tau
_{1}b_{x}e^{i\int_{0}^{\tau _{1}}d\tau _{2}b_{l}(\tau _{2})}\right| }.
\end{equation}
This result is still valid for any time-dependent external fields. We give
two examples of special time-dependent external magnetic fields.

(a) Sinusoidal field $b_{l}(t)=b\cos (\omega t)$. According to Eq. (18), the
oscillation period is

\ 
\begin{equation}
T=\frac \pi {b_xJ_0(b/\omega )},
\end{equation}
where $J_0$ is the zeroth order Bessel function of the first kind. From this
result, it can be seen that when $J_0(b/\omega )\neq 0$, there exists a
finite oscillation period, indicating that the population can transit from
an initial polar state to other spin states within the driving process. When 
$b=0$, the longitudinal field-free oscillation has period $T_0=\pi /b_x$.
The fact $\left| J_0(b/\omega )\right| \leq 1$ implies that $T\geq T_0$.
This shows that the invasion of the longitudinal magnetic field will
suppress the coherent population transfer among the five spin components.
The extreme case occurs when the reduced amplitude of the longitudinal
magnetic field $b/\omega $ is a root of $J_0$. In this case, we have $%
T\rightarrow \infty $, which implies that the population transfer is totally
suppressed. Hence the system will stay in the initial polar state during the
whole driving process. This is just the phenomenon of spin localization. The
above analysis is illustrated in Fig. 1, where the scaled oscillation period 
$T/T_0$ is plotted against $b/\omega $. It can be seen that the oscillation
period increases significantly when $b/\omega $ approaches the roots of
Bessel function $J_0$.

(b) A combination of static and sinusoidal magnetic fields $%
b_{l}(t)=b_{0}+b\cos (\omega t)$. In this case, the oscillation period is

\begin{equation}
T=\frac{\pi }{\left| b_{x}\lim_{N\rightarrow \infty }\frac{\sin (b_{0}\pi
N/\omega )}{N\sin (b_{0}\pi /\omega )}\widehat{J}_{b_{0}/\omega }(b/\omega
)\right| },
\end{equation}
where $\widehat{J}$ is the Anger function\cite{ref10} defined by

\begin{equation}
\widehat{J}_{a}(b)=\frac{1}{\pi }\int_{0}^{\pi }dx\cos (ax-b\sin x).
\end{equation}
There exists an infinite oscillation period when $b_{0}/\omega \neq k$ ($k$
is an integer), which means that the population time evolution will undergo
localization in the initially populated spin-0 state during the whole
driving process. When $b_{0}/\omega =k$, the oscillation period becomes

\begin{equation}
T=\frac \pi {b_xJ_k(b/\omega )},
\end{equation}
where $J_k$ are $k$-th order Bessel functions. From this result, we find
that we can have a finite oscillation period only when $b/\omega $ is not
equal to a root of $J_k$, and in this case the evolution of the initial
polar phase is delocalized into other spin states. If $b/\omega $ becomes a
root of $J_k$, the oscillation period approaches infinity, and we have the
spin localization.

\section{EFFECTS OF EXTERNAL NOISE}

In practice, since the external magnetic field may have a fluctuation
component, the effects of external noise has to be considered. Another
motivation for the introduction of an external noise is to provide the
fluctuations required to destroy the coherence of the population transfer
process and, in suitable cases, leads to a rate process for the decay of the
population. In the following, we study the dynamics of the system when the
external magnetic field has a fluctuating component. In the simplest case,
the statistical properties of the imposed magnetic field can be prescribed
to be independent of the characteristics of the system. As a versatile
choice for the noise, an Ornstein-Uhlenbeck (OU) process\cite{ref11} is
used. This permits the investigation of the role of the strength and size of
correlation time of the noise on the time evolution of the spinor BEC.
Without loss of generality, the external longitudinal magnetic field $b_l(t)$
is assumed to contain two components, a stochastic part $f(t)$ and a
systematic part $b_0+b_1\cos (\omega t)$, i.e.,

\begin{equation}
b_{l}(t)=b_{0}+b_{1}\cos (\omega t)+f(t).
\end{equation}
The noise $f(t)$ is assumed to be characterized by an OU process whereby it
has zero average value and correlation function

\begin{equation}
\langle f(t)f(s)\rangle =\Delta ^2\exp (-\left| t-s\right| /\tau _c).
\end{equation}
Here the quantities $\Delta $ and $\tau _c$ are the strength and decay
constant of the noise, respectively. When the noise is external to the
system, and therefore not necessarily thermal in character, $\Delta $ and $%
\tau _c$ can be varied in a controlled manner, and are not restricted by the
physical properties of the system. In the absence of an analytic solution, a
numerical investigation of population $P_\alpha (t)$ is adapted. The
population distribution in the five components of spinor BEC with the
external noise are solved numerically by generating trajectories for the
different realization of the noise. The procedure\cite{ref12} to integrate
the stochastic equations is as follows. A stochastic term is added at each
step with its statistical properties described by an OU process. The OU
process is generated by solving a Langevin equation with a delta-correlated
noise term. This ensures that the correlation function of $f(t)$ has the
desired statistical property given by Eq. (24).

We present in Fig. 2 the time evolution of the population $P_0(t)$ of spin-0
state in the static magnetic field [$b_1=0$ in Eq. (23)] for the value of $%
b_0=10b_x$. In the absence of noise, as shown in Fig. 2(a), the population
of spin-0 state is always close to unity during time evolution. This effect
is purely the static spin localization. In this case, the longitudinal
magnetic field lifts the degeneracy of the five spin components and
consequent energy mismatch hinders the BEC transfer from the initially
occupied polar state to its neighbours in the spin space. In the case of the
intermediate ($\Delta \tau _c\sim 1$) noise modulation, as shown in Fig.
2(b) (solid curve), it can be seen that the population of spin-0 state has a
damped oscillation with a slow monotonic decay superimposed. The combination
of noise and static magnetic field, suppressing population transfer,
produces the decay. It also shows in Fig. 2(b) that with increasing strength
of noise, the decay becomes more rapid. Therefore, the decaying time can be
controlled by varying the magnetic field or the strength of noise.

Figure 3 shows the time dependencies of $P_0(t)$ in the time-dependent
longitudinal magnetic field with the systematic part $b_l=b\cos (\omega t)$.
The field parameters are chosen as $b=2b_x$ and $\omega =10b_x$,
corresponding to the Rabi oscillation with scaled period $T=T_0/J_0(0.2)$,
as shown in Fig. 3(a). In the case of a weak noise $\Delta /b_x=0.2$ (dashed
curve), the population $P_0$ is still oscillatory in our scope of time, but
its oscillation amplitude decreases with time. When the strength of the
noise increases to $\Delta /b_x=1.0$ (dotted curve), the Rabi oscillation
breaks down completely and the system decays fast towards equilibrium. This
suggests that the coherent Rabi oscillation for a spinor BEC is sensitive to
the external noise and is destroyed even in a weak coupling regime. We also
present in Fig. 4 the time evolution of $P_0$ for the value of $b/\omega
=2.402$, corresponding to the dynamic localization, as shown in Fig. 4(a).
In Fig. 4(b), we can see that in the case of intermediate ($\Delta \tau
_c\sim 1$) noise modulation (solid curve), the population $P_0$ still
remains close to unity for an extremely long time. This insensitivity to the
presence of weak noise reflects the strength of the systematic field. When
the strength of noise increases to a strong coupling regime, a more rapid
and less oscillatory decay is observed in the population evolution, as shown
in Fig. 4(b) (dotted curve).

Figure 5 shows the time evolution of $P_{0}$ in the external longitudinal
magnetic fields consisting of static and time-periodic components [Eq.
(23)]. The field parameters are chosen as $b_{0}/b_{x}=10$, $\omega
/b_{x}=10 $, and $b_{1}/b_{x}=38.3$, corresponding to the spin dynamic
localization, as shown in Fig. 5(a). In the case of intermediate ($\Delta
\tau _{c}\sim 1$) noise modulation, similar to that shown in Fig. 4(b), the
population $P_{0} $ of spin-0 state remains close to unity in our scope of
time. Even in the strong coupling regime (dashed curve), the fast
oscillations still exist, suggesting the fundamental interplay between the
systematic and noise fields. With further increase of the strength of noise,
the system decays rapidly towards equilibrium, and therefore, the spinor BEC
initially localized in spin-0 state is delocalized and diffuses among all
spin states, as shown in Fig. 5(b) (dotted curve).

\section{Dynamic spin localization beyond single-mode approximation}

In the above discussions, the single-mode assumption (SMA) has been used:
the wave function for each spin component retains the same spatial profile
during the time development. Therefore, the possibility of high-energy mode
excitation is neglected and the system dynamics is fully described by the
internal population transfer among the spin components. This SMA treatment
is valid for a small particle number $N$. When $N$ is large, the nonlinear
spin-mixing processes induce the large energy transfer from the initial
ground state to the excited modes, thus destroying the validity of the SMA.
However, when the initial state is taken to be the eigenstate of the
mean-field G-P equation, the SMA is expected to describe well the dynamics
in the absence of dissipation and external driving\cite{Pu2}. In this
section, we go beyond the SMA by means of a numerical simulation of the
time-spatial evolution of the system.

Due to anisotropic nature of the optical trapping potential, the cigar
shaped BEC is assumed to be quasi-one-dimensional and hence the wave
function associated with the spin-$\alpha $ state may be written as 
\begin{equation}
\psi _\alpha (x,y,z,t)=\phi _{\bot }(x,y)\phi _\alpha (z,t)e^{-i\omega
_{\bot }t},
\end{equation}
where $z$ is the direction of weak confinement, $\phi _{\bot }(x,y)$ is the
ground-state wave function of the two-dimensional harmonic oscillator, and $%
\omega _{\bot }$ the tight confinement frequency. Inserting Eq. (25) into
Eqs. (1) and (2), we obtain the equations of motion for the longitudinal
wave functions $\phi _\alpha (z,t)$ in a dimensionless form 
\begin{equation}
i\hslash \frac{\partial \psi _2}{\partial t}={\cal L}\psi _2+f_{-}\psi
_1+f_z\psi _2+\psi _{-2}^{*}s_{-}-b_x\psi _1-2b_l(t)\psi _2  \eqnum{26a}
\end{equation}
\begin{equation}
i\hslash \frac{\partial \psi _1}{\partial t}={\cal L}\psi _1+f_{+}\psi _2+%
\sqrt{\frac 32}f_{-}\psi _0+f_z\psi _1-\psi _{-1}^{*}s_{-}-b_x(\psi _1+\sqrt{%
\frac 32}\psi _0)-b_l(t)\psi _1  \eqnum{26b}
\end{equation}
\begin{equation}
i\hslash \frac{\partial \psi _0}{\partial t}={\cal L}\psi _0+\sqrt{\frac 32}%
(f_{+}\psi _1+f_{-}\psi _{-1})+\psi _0^{*}s_{-}-b_x(\psi _1+\psi _{-1}) 
\eqnum{26c}
\end{equation}
\begin{equation}
i\hslash \frac{\partial \psi _{-1}}{\partial t}={\cal L}\psi _{-1}+\sqrt{%
\frac 32}f_{+}\psi _0+f_{-}\psi _{-2}-f_z\psi _{-1}-\psi _1^{*}s_{-}-b_x(%
\sqrt{\frac 32}\psi _0+\psi _{-2})+b_l(t)\psi _{-1}  \eqnum{26d}
\end{equation}
\begin{equation}
i\hslash \frac{\partial \psi _{-2}}{\partial t}={\cal L}\psi _2+f_{-}\psi
_{-1}-f_z\psi _{-2}+\psi _2^{*}s_{-}-b_x\psi _{-1}+2b_l(t)\psi _{-2} 
\eqnum{26e}
\end{equation}
where ${\cal L}=-d^2/dz^2+z^2/4+\overline{c}_0N\eta (|\phi _2|^2+|\phi
_1|^2+|\phi _0|^2+|\phi _{-1}|^2+|\phi _{-2}|^2)$ with transverse structure
factor $\eta =\int |\phi _{\bot }(x,y)|^4dxdy/\int |\phi _{\bot
}(x,y)|^2dxdy $, $f_{+}=\overline{c}_1(\phi _2^{*}\phi _1+\sqrt{3/2}\phi
_1^{*}\phi _0+\sqrt{3/2}\phi _0^{*}\phi _{-1}+\phi _{-1}^{*}\phi _{-2})$, $%
f_{-}=f_{+}^{*}$, $f_z=(\overline{c}_1/2)\sum_{i=-2}^2|\phi _i|^2$, and $%
s_{-}=(2\overline{c}_2/5)(\phi _0^2/2-\phi _1\phi _{-1}+\phi _2\phi _{-2})$.
The above equations have been written in dimensionless form and the units
for length, energy, and time are $\sqrt{\hslash /(2m\omega _z)}$, $\hslash
\omega _z$, and $1/\omega _z$, respectively, where $\omega _z$ is the axial
trapping frequency. In the numerical simulation of the dynamic equations
(26), the initial wave functions are taken to be ground state of the
one-dimension G-P equations ${\cal L}\phi _i(z,0)=\mu \phi _i(z,0)$, where $%
\phi _i(z,0)$ satisfies the normalization $\int |\phi _i(z,0)|^2dz=N_i/N$
with $N_i$ the initial particle number in spin-$i$ component.

We first discuss the dynamics of the system in the absence of coupling
magnetic fields. Figure 6 shows the time evolution of the population of spin-%
$0$ component for two kinds of initial population. The solid line in Fig. 6
corresponds to the initial state that five spin components are initially
equally populated, whereas the dotted line corresponds to the case that the
BEC initially populates spin-0 components. It shows (solid line) that during
time evolution, the population of spin-0 component develops into a complex
oscillatory structure, implying the excitation of high-energy modes. The
excitation originates from the fact that the initial state is not a
mean-field eigenstate of the SMA Hamiltonian (6), thus leading to a
redistribution of the total energy $E$ between the symmetric and
unsymmetrical parts (denoted by $E_{s}$ and $E_{a}$, respectively) of the
total Hamiltonian. When $E_{a}$ is sufficiently large, the higher modes are
excited by the nonlinear spin-mixing processes and the SMA description
breaks down for large value of particle number $N$. When the initial state
is chosen to be the SMA eigenstate, the population of spin-$0$ state does
not vary in time (dotted line). Thus, SMA treatment is appropriate for the
spin polarized state with the BEC localized in spin-0 component.

In the presence of external magnetic fields, the five spin components are
coupled. The atomic population is featured by Rabi oscillations among the
spin states. To see the validity of the SMA in studying the dynamic spin
localization, We present in Fig. 7 the time evolution of the population of
spin-0 state for several values of amplitude of oscillatory magnetic field.
It shows that the weak transverse magnetic field couples the five spin
components, inducing Rabi oscillation between the initial populated spin-0
state and the other spin states. The presence of an additional oscillatory
magnetic field slows down the oscillation periods. Finally, when the field
parameter is modulated to satisfy $b/\hslash \omega =2.4$, as shown in Fig.
7(b), the Rabi oscillation completely stops and population of spin-0 state
always remains close to unity during its time development. The good
agreement between the SMA results and exact numerical solution suggests that 
{\it the phenomenon of dynamic spin localization is well-defined beyond the
simple SMA.} This robustness of dynamic localization against nonlinear
atomic interaction is contrast to the case of two-component BEC system,
where there is no energy-stable spin-polarized states.

\section{CONCLUSION}

In summary, by the application of the external magnetic fields which may
contain a random component, we have elaborated on the control of population
transfer among the internal spin states for the $F=2$ spinor BEC. After
taking the mean-field approximation a general formula is obtained for the
oscillation period to describe the quantum transition from the initial spin
polar state to other spin states. In particular, when the frequency and
reduced amplitude of the longitudinal magnetic field are related in a
specific manner, the population of the initial spin-0 state is dynamically
localized during the time evolution. In addition, the effects of external
noise on the coherent population transfer dynamics of the spinor BEC are
studied. We found that (i) The Rabi oscillations are sensitive to the
external noise, and even weak coupling can destroy completely the Rabi
oscillations in an earlier length of time. (ii) When the strength of noise
is weak, because of strong suppression caused by the systematic field, the
static and dynamic localization may remain for experimentally relevant
times. (iii) When the strength of noise is strong, \ delocalization due to
phase randomness occurs during time evolution and the initially localized
BEC diffuses among five spin components. (iv) The Rabi oscillation and
dynamic spin localization is robust against the nonlinear spin exchange
interaction. Thus in order to observe the static or dynamic localization in
spin space one should produce stable magnetic fields, which has as little as
possible the stochastic component. Although it is difficult to keep the
external noise from intruding on dynamics, recent observation of the
well-defined coherence between the two BECs\cite{Hall,Andrews} suggests that
the external noise can be greatly suppressed in experimental setup.
Therefore, we expect such macroscopic coherent quantum controls as Rabi
oscillation and dynamic localization may be realized in spinor BEC
experiments.

{\large ACKNOWLEDGMENT}

This work was partially supported by a grant from the Research Committee of
The Hong Kong Polytechnic University (Grant No. G-T308).

{\Large Figure captions}

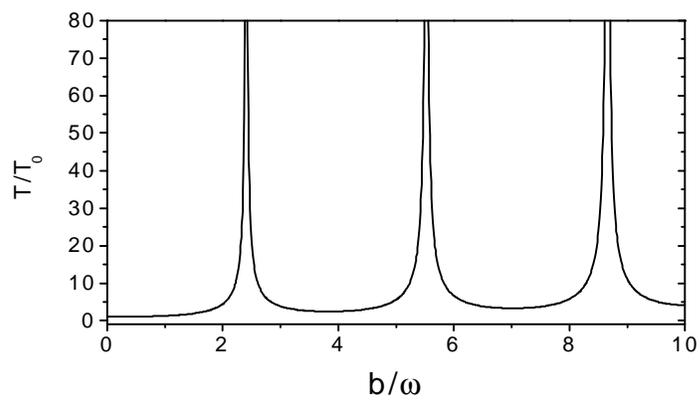
Fig. 1. Scaled oscilltion period $T/T_0$ of the sinusoidal magnetic field
case, versus the reduced field strength $b/\omega $.

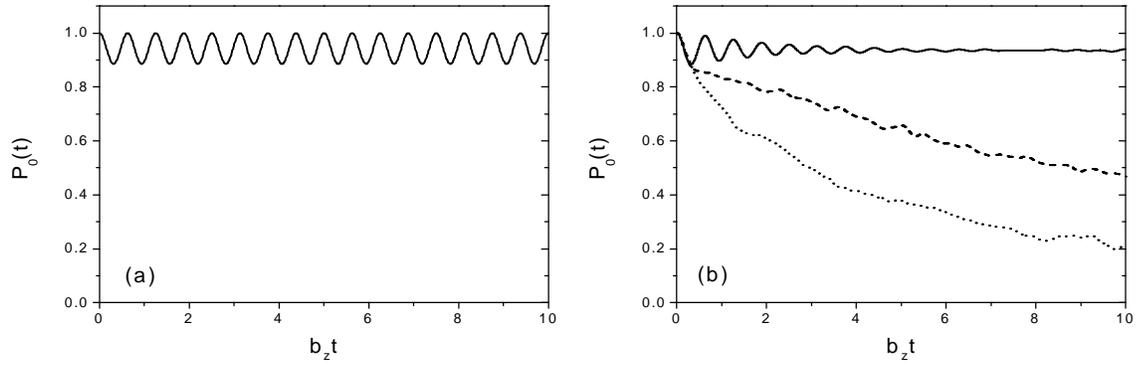
Fig. 2. Time evolution of $P_0(t)$ with and without external noise in the
static magnetic field, $b/b_x=10$. (a) the case without the noise, (b) the
case with the noise $\Delta /b_x=1.0$ (solid curve), $\Delta /b_x=5.0$
(dashed curve), and $\Delta /b_x=10.0$ (dotted curve). $b_x\tau _c$ is fixed
to be 1.0 for the case with the noise.

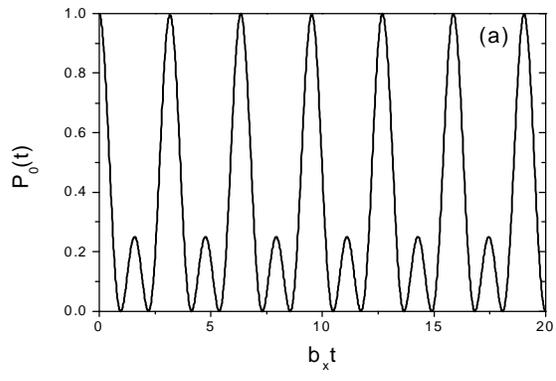
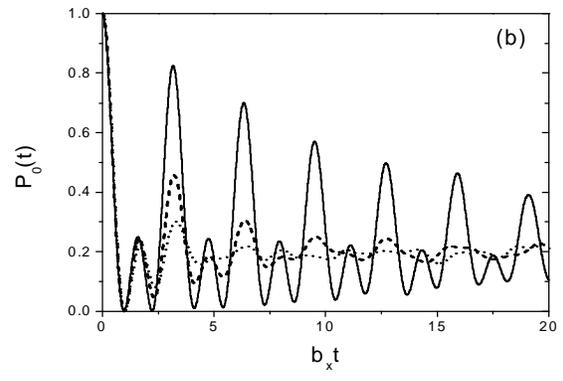
Fig. 3. Time evolution of $P_0(t)$ with and without external noise in
sinusoidal magnetic field, $b/b_x=2$, $\omega /b_x=10$. (a) the case without
the noise, (b) the case with the noise $\Delta /b_x=0.2$ (solid curve), $%
\Delta /b_x=0.6$ (dashed curve), and $\Delta /b_x=1.0$ (dotted curve). $%
b_x\tau _c$ is fixed to be 1.0 for the case with the noise.

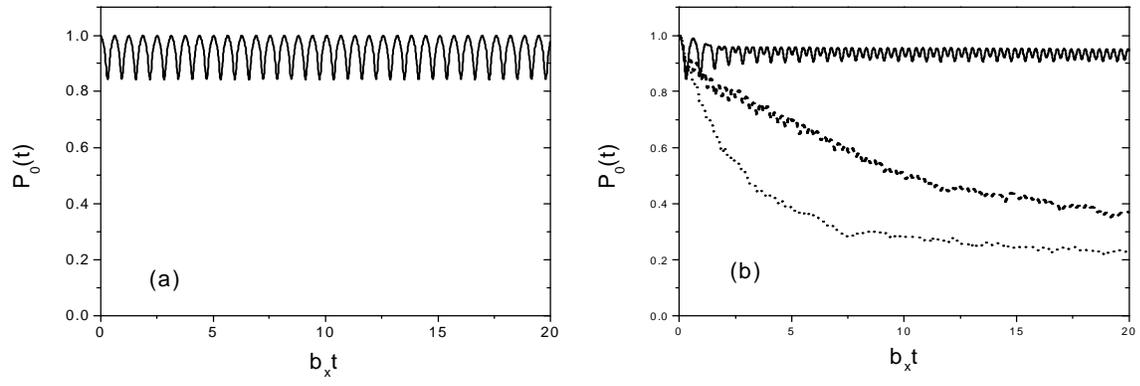
Fig. 4. Time evolution of $P_0(t)$ with and without external noise in
sinusoidal magnetic field, $b/b_x=24.02$, $\omega /b_x=10$. (a) the case
without the noise, (b) the case with the noise $\Delta /b_x=1.0$ (solid
curve), $\Delta /b_x=5.0$ (dashed curve), and $\Delta /b_x=10.0$ (dotted
curve). $b_x\tau _c$ is fixed to be 1.0 for the case with the noise.

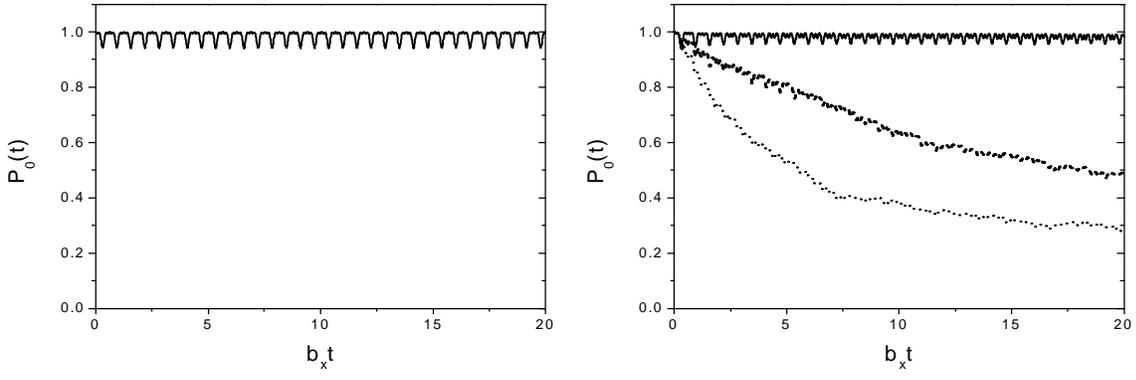
Fig. 5. Time evolution of $P_0(t)$ with and without external noise in a
combination of static and sinusoidal magnetic field, $b_0/b_x=10$, $%
b_1/b_x=38.3$, $\omega /b_x=10$. (a) the case without the noise, (b) the
case with the noise $\Delta /b_x=1.0$ (solid curve), $\Delta /b_x=5.0$
(dashed curve), and $\Delta /b_x=10.0$ (dotted curve). $b_x\tau _c$ is fixed
to be 1.0 for the case with the noise.

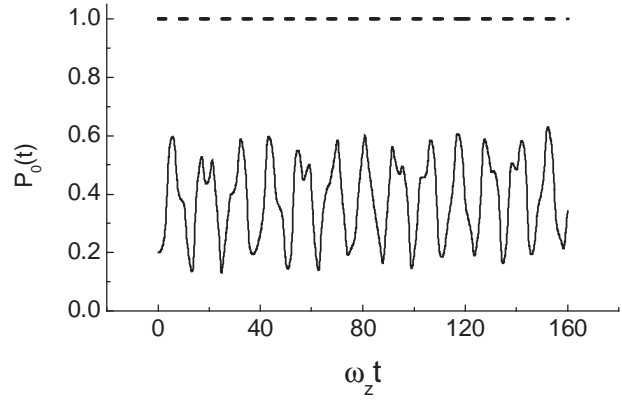
Fig. 6. time evolution of $P_{0}(t)$ for two kinds of initial population: $%
N_{i}/N=0.2,$ $i=-2,...,2$ (solid line), $N_{0}/N=1$ (dashed line). Other
parameters are $\omega _{z}=2\pi \times 60$Hz, $a_{0}=34.9a_{B}$, $%
a_{2}=45.8a_{B}$, $a_{4}=64.5a_{B}$ for $^{23}$Na ($a_{B}$ is the Bohr
radius), $\eta =1$, $N=10000$.

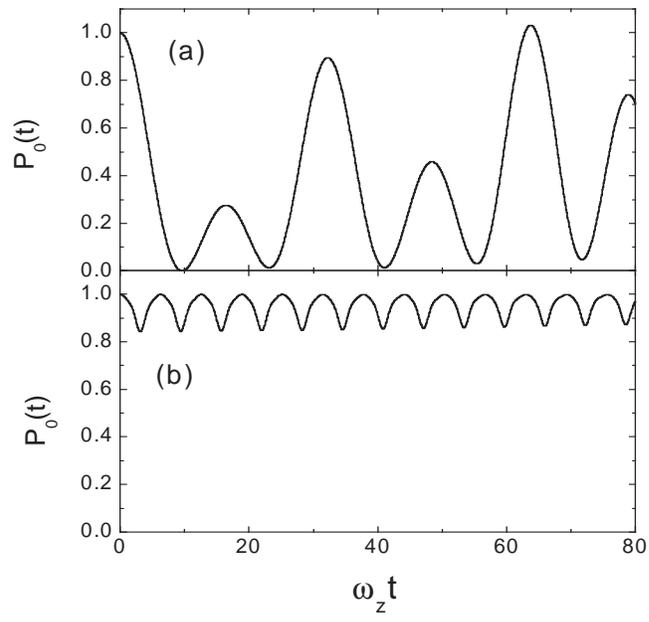
Fig. 7. Time evolution of $P_{0}(t)$ in the presence of a uniform field $%
b_{x}=1$. (a) the case without sinusoidal magnetic field, (b) the case with
sinusoidal magnetic field $b_{l}(t)=b\cos (\omega t)$, $\omega =10$, $b=24$.
Other parameters are the same as used in Fig. 6.

\end{document}